\documentclass[apjl]{emulateapj}
\usepackage{graphicx}
\usepackage{amssymb}
\usepackage{amsmath}

\def\cxo{{\sl CXO\ }}

\def\gtrsim{\mathrel{\hbox{\rlap{\hbox{\lower4pt\hbox{$\sim$}}}\hbox{$>$}}}}
\def\lesssim{\mathrel{\hbox{\rlap{\hbox{\lower4pt\hbox{$\sim$}}}\hbox{$<$}}}}
\def\gtrsim{\mathrel{\hbox{\rlap{\hbox{\lower4pt\hbox{$\sim$}}}\hbox{$>$}}}}
\def\farcs{\hbox{$.\!\!^{\prime\prime}$}}

\begin{document}

\title{
An extended X-ray 
object ejected from the PSR\,B1259--63/LS\,2883 binary}

\author{
George G. Pavlov\altaffilmark{1},
Jeremy Hare\altaffilmark{2},
Oleg Kargaltsev\altaffilmark{2},
Blagoy Rangelov\altaffilmark{2},
Martin Durant\altaffilmark{3}
}

\altaffiltext{1}
{Department of Astronomy \& Astrophysics, Pennsylvania State University, 525 Davey Lab, University Park, PA 16802, USA; ggp1@psu.edu}
\altaffiltext{2}
{George Washington University, 105 Corcoran Hall, Washington, DC 20052, USA}
\altaffiltext{3}
{Toronto, ON, Canada}

\begin{abstract}
We present the analysis of the  {\sl Chandra X-ray Observatory} observations of the eccentric $\gamma$-ray binary PSR\,B1259--63/LS\,2883. 
The analysis shows that the
 extended X-ray  feature seen in previous observations
is still moving away from the binary with an 
average 
projected velocity of $\approx0.07c$ and shows a hint of acceleration. The spectrum of the feature appears to
 be hard (photon index $\Gamma\approx 0.8$) with no sign of softening compared to previously measured values.
We interpret it as a clump of plasma ejected from the binary through the interaction of the pulsar with the decretion disk
 of the O-star around 
periastron passage.
 We suggest that the clump is 
 moving in the unshocked
relativistic pulsar wind (PW), which can accelerate the clump.  
Its X-ray emission can be interpreted as synchrotron radiation of
the PW shocked by the collision with the clump.
\end{abstract}


\section{Introduction}

High-mass $\gamma$-ray binaries (HMGBs)
consist of a compact object (i.e., a black hole or a neutron star) 
and a massive (early-B or late-O type) star. Only one of the few
known HMGBs, B1259--63/LS\,2883 (hereafter B1259), has a compact object that has been detected as a radio pulsar. The pulsar has the following properties: period {\sl P}=47.8 ms, characteristic age $\tau=P/2\dot{P}=330$  kyr, spin-down power $\dot{E}$=8.3$\times 10^{35}$ erg  s$^{-1}$, magnetic field $B$ $= 3.3\times 10^{11}$ G, and distance $d$ $\approx$ 2.3 kpc \citep{1992ApJ...387L..37J}. 
The pulsar's companion is a fast-rotating O star, with
mass $M_*\approx 30 M_\sun$ and
 luminosity $L_{*}=6.3 \times 10^{4}L_{\sun}$ \citep{2011ApJ...732L..11N},
and an equatorial 
decretion disk\footnote{ Decretion disks (also known
as excretion disks) are viscous gaseous disks around  rapidly rotating
O or B stars (Be stars) in which stellar matter slowly moves outward and rotates with
nearly Keplerian velocity in the equatorial plane (see Rivinius et al.\ 2013 for a recent review).}, which is inclined at an angle of 
$\approx 35^\circ$ to the orbital plane \citep{1995MNRAS.275..381M, 2014MNRAS.437.3255S}.
The system has an orbital period  $P_{\text{orb}}= 1236.7$ days, an eccentricity $e=0.87$, a semi-major axis  $a\approx 7$AU, and an inclination angle $i\approx 23^{\circ}$ \citep{2011ApJ...732L..11N}.
The pulsar is eclipsed for 
about one month around periastron \citep{2005MNRAS.358.1069J}
 when it is moving through the 
decretion disk and dense wind of its companion star.
During the pulsar eclipse, peaks of nonpulsed radio emission
were observed,
centered at $\approx 5$ days before periastron and $\approx 20$ days after periastron, with the
relative strengths of the peaks varying from cycle to cycle.
\cite{1999ApJ...514L...39B}
 interpreted it is as synchrotron radiation of electrons 
accelerated during the pulsar passages through the decretion disk.

B1259 has been extensively studied in X-rays and $\gamma$-rays.
The X-ray emission from B1259 showed no pulsations. It was interpreted as synchrotron radiation from
relativistic pulsar wind (PW) leptons accelerated at the shock between the PW
and the massive star outflow \citep{1997ApJ...477..439T}.
Multi-epoch measurements with numerous X-ray observatories
revealed binary phase dependences of 
the flux, photon index, 
and absorption column density 
(\citeauthor{2006MNRAS.367.1201C} \citeyear{2006MNRAS.367.1201C}, 
\citeyear{2009MNRAS.397.2123C}).
Within the same orbital cycle, the strongest variations are seen around periastron.
In particular, the X-ray light curve shows two peaks corresponding to 
first and second passages of the pulsar through the disk
($\approx 20 - 4$ days before periastron and 
$\approx 10 - 50$ days after periastron), while the
hydrogen column density, $N_H \approx 3\times 10^{21}$ cm$^{-2}$ during most
of the cycle, increases 
by a factor of 2 during and between the two disk passages,
confirming the presence of additional absorbing matter at these binary phases.

This unique system is one of the few HMGBs whose flaring in the GeV energy range was seen by the {\sl Fermi} Large Area Telescope.
The GeV flares of B1259
were detected
soon after the pulsar passed through periastron in 2010 and 2014,
while no GeV emission was seen when it was far from periastron
\citep{2011ApJ...736L..11A, 2011ApJ...736L..10T,  
2015ApJ...798L..26T}.
The flares started $\approx 20 - 25$ days after the epochs of periastron and
lasted roughly 1 -- 2 months.
This GeV emission could be synchrotron radiation from PW electrons 
accelerated to energies $\sim 100$ TeV in the intra-binary shock,
or inverse Compton (IC) scattering of photons emitted by the
massive companion off electrons with energies $\sim 1 - 10$ GeV
\citep{2011ApJ...736L..11A}.

In the TeV energy range, B1259 has been observed by the H.E.S.S.\ observatory
\citep{2005A&A...442....1A, 2009A&A...507..389A}.
Variable TeV emission 
was detected around the periastron passages of 2004 and 2007, 
with light curves similar to those of the
nonpulsed radio emission.
No signatures of TeV emission enhancement were detected at the times of GeV
flares after
the 2010 and 2014 periastron passages
(\citeauthor{2011ApJ...736L..11A} \citeyear{2011ApJ...736L..11A}; \citeauthor{2014ATel.6204....1M} \citeyear{2014ATel.6204....1M}).
Several models have been developed to explain the TeV emission, such as 
IC scattering of stellar photons 
off ultra-relativistic electrons (\citeauthor{2009A&A...507..389A} \citeyear{2009A&A...507..389A}; \citeauthor{2007MNRAS.380..320K} \citeyear{2007MNRAS.380..320K}) and 
hadronic interaction of the PW with the dense decretion disk \citep{2009A&A...507..389A}.

The size of the B1259 binary is too small to resolve it in X-rays, but the high
angular resolution of the {\sl Chandra X-ray Observatory} ({\sl CXO}) allows
one to look for extended emission from the pulsar wind nebula (PWN)
or from matter that could be expelled from the binary. First detection of
extended emission associated with B1259 was reported by
\cite{2011ApJ...730....2P}.
They observed B1259 with {\sl CXO} on 2009 May 14 (ObsID 1089; 26 ks exposure),
when the pulsar just had passed the apastron of 2009 April 5
(see Figure 1), and detected faint,
asymmetric extended emission seen up to $\approx 4''$ south-southwest from
the binary position.
This emission was interpreted as 
a PWN blown out of the binary by the massive companion's wind.

\begin{figure}[ht!]
\centering
\includegraphics[width=8cm]{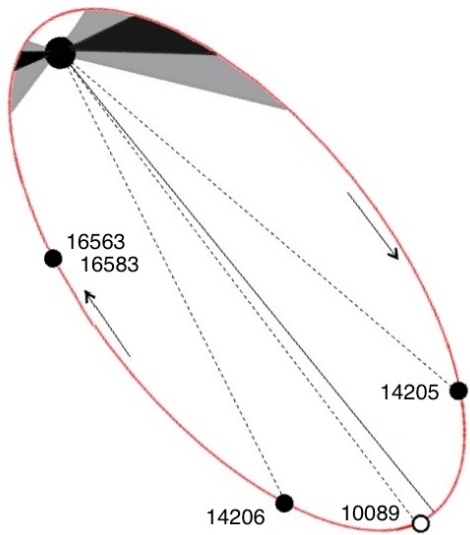}
\vspace{-1cm}
\caption{
Orbital locations of the pulsar 
at the times of 
our {\sl Chandra} ACIS observations, marked by their ObsIDs. Our `new data' consists of two exposures (ObsIDs 16563 and 16583) separated by about 10 hours. Gray areas show the parts of the orbit where the pulsar passes through the companion's equatorial disk.
\label{pic1}
}
\end{figure}

\citeauthor{2014ApJ...784..124K} (\citeyear{2014ApJ...784..124K}; 
referred to as 
K+14 hereafter)
reported results  of two deeper
{\sl CXO} observations 
(ObsIDs 14205 and 14206; 56 ks exposures).
These observations were carried out on 2011 December 17 and 2013 May 19,
i.e., 
370 and 886 days after the periastron of 2010 December 14.
They revealed a variable, extended
(about $4''$, $\sim1000$ times the binary orbit size) structure, which appeared to shift by $1\farcs8\pm 0\farcs5$ between the two observations. K+14 discussed
 two possible interpretations of the observed variability: (1) an extrabinary
termination shock 
in the PW outflow, with a variable stand-off distance,
 or  (2) a clump ejected from the binary and moving with a projected velocity 
$v_\perp  =(0.046 \pm 0.013)c$, at $d=2.3$ kpc. The analysis of 
those data favored the scenario in which most of the observed extended emission would come from the PW launched from the binary near the apastron of 2009. To gain a better understanding of the observed phenomenon and test our interpretations,
we proposed a Director Discretionary Time (DDT) {\sl CXO} observation. 
Here we report the results of the new observation, analyze it jointly with the 
two previous observations (taken within the same orbital cycle between the
periastrons of 2010 and 2014), and suggest a new interpretation of the
extended structure.

\section{Observations and Data Reduction}

B1259 was observed with the Advanced CCD Imaging Spectrometer (ACIS) 
on-board {\sl CXO} on 2014 February 8 and 9 (ObsIDs 16563 and 16583, respectively),
with a total live time of 57.6 ks.
The corresponding orbital location of the pulsar ($\approx 1150$ days after
the peristron of 2010; $\theta=221^\circ$)
 is shown in Figure 1.
The observation setup was similar to our previous observations that have been analyzed in K+14.
The target was imaged on the front-illuminated ACIS-I3 chip in timed exposure mode,
the data were telemetered in `very faint' format. A 1/8 subarray was used to reduce the frame time to 0.4 s and mitigate the effect of pile-up. The highest count rate observed from the binary (0.043 counts per frame) corresponds to a negligibly small pile-up fraction of $< 2\%$.

We used the pipeline-produced Level 2 event files for the analysis. No episodes of anomalously high background rates occurred in the observations.  To
minimize the background contribution, we filtered the events by limiting the photon energies to $0.5-8 $ keV.
The detector response for spectral analysis was produced with CIAO (ver.\ 4.6) tools, following the standard procedure and using the calibration database CALDB 4.5.9. The spectral fitting was done with XSPEC (ver.\ 12.8.0).

\section{Data Analysis }

\subsection{
Apparent motion of the extended feature}
\begin{figure*}[ht!]
\centering
\includegraphics[width=\textwidth]{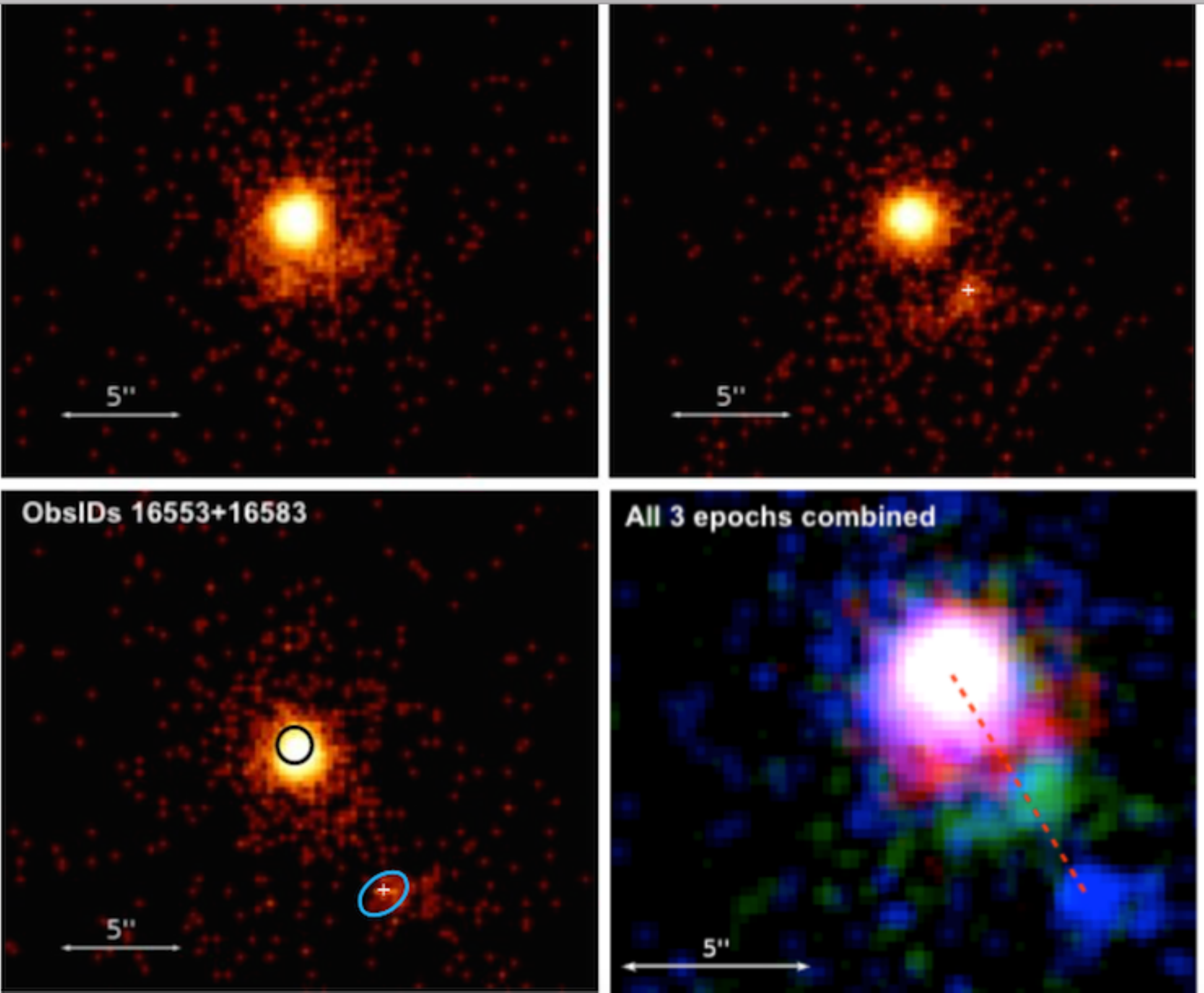}
\caption{ 
ACIS images from our previous two observations, ObIDs 14205 (top left) and 14206 (top right), and the new data (bottom left). The blob of extended emission is clearly traveling away from the binary. The bottom right panel shows the false color
image (different colors correspond to different observation epochs) demonstrating the motion and changes in morphology. The dashed line connects the binary and the centroid of the feature in the new data image.  The angular separation between the extended feature and the binary  as a function of time is shown in Figure \ref{pic2}.  The spectral extraction regions (core and extended emission are shown in black and cyan,
 respectively) on 
new data image. North is up, east is to the left. The radius of the black circle is 0\farcs9; the background extraction region (not shown) is an annulus with radii $8\farcs55$ $\leq r \leq $ $19\farcs18$. The white crosses mark the positions of the extended emission 
as obtained with the CIAO {\tt celldetect} tool. The apparent asymmetry of the bight core seen in the merged image of ObsIDs 16563+16583 can, at least partly, be attributed to the known 
Chandra mirror asymmetry.
 \label{pic8}\label{images}}
\end{figure*}

Since the two new observations
were carried out less than one day apart, we merged the data for purposes of
image
analysis.
The merged image clearly reveals a `blob' of extended emission with a very  faint westward extension. The appearance of the extended feature has changed substantially between the preceding observation ObsID 14206 and the new data (see Figure  \ref{pic8}). Unlike the earlier images, in which the feature resembled an arc, the feature's shape is more round in the new image.  It is possible that the fainter part of the feature had simply become too faint to be detected as a result of a
decrease in the surface brightness.

\begin{figure}[ht!]
\includegraphics[width=8.5 cm]{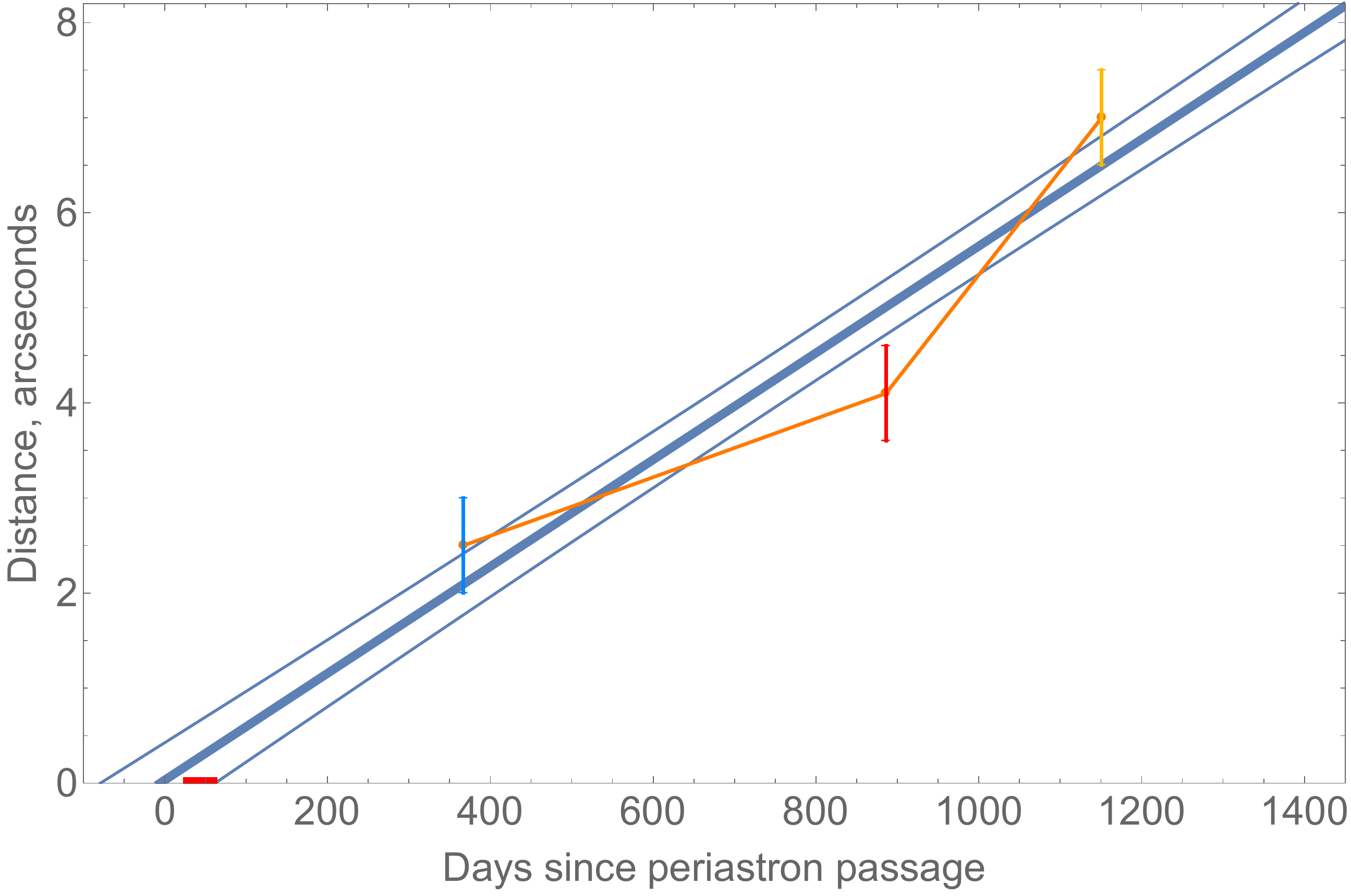}
\caption{
Separation of the extended 
feature from the unresolved point source (in arcseconds) as a function of time since the periastron passage on 2010 December 14. The line of best fit and  $1\sigma$ 
upper and lower bounds are also shown. The small red line at the bottom shows the time period, 2011 January 14 -- March 4, in which the GeV flare
occurred 
\citep{2011ApJ...736L..11A}. 
\label{pic2}
}
\end{figure}

The new data allow us to 
further track the motion of the extended feature reported in K+14. Using the 
CIAO {\tt celldetect} tool,
we measured the positions of the binary and the extended feature together with the corresponding  $1\sigma$ uncertainties\footnote{
For the elongated arc-like structure in the earlier images, K+14 used a different method (see Figure 3 in K+14), 
but running {\tt celldetect} on ObsID 14206 gave consistent results. { For 
 ObsID 14205 {\tt celldetect} fails to pick out the azimuthally extended 
structure because it was too close to the bright binary, so we used 
the peak in the radial count distribution for the blob position and the
FWHM of the peak for its uncertainty.}}. 
The $3\farcs1 \pm 
0\farcs7$ shift of the blob, with respect to the position  of the arc-like structure in the ObsID 14206 image, corresponds to a proper motion 
$\mu = 4\farcs2 \pm 0\farcs9$ yr$^{-1}$ and projected velocity 
$v_{\perp}=
(0.15\pm0.04)c$,  at $d= 2.3$ kpc. 
The temporal dependence of the
 radial separation of the extended emission from the binary, 
at a position angle of $215^\circ$ (north through east),
is plotted in Figure \ref{pic2}. The plot does not show any evidence of 
deceleration, rather it might suggest that the blob is moving 
with an acceleration
$\dot{v}_\perp \sim \Delta v_\perp/\Delta t \sim (0.010\pm 0.04) c/(390\,{\rm d})=
90\pm 40$ cm\,s$^{-2}$.
Given the low significance of this result, we assumed a constant
velocity and estimated it
by fitting the radial separations with a straight line, $r(t)=
\mu (t-t_0)+r_0$, where $t_0=801$ days is the reference time (all times are counted from the periastron of 2010 December 14). The fit yields $r_0=4\farcs6\pm 0\farcs4$, $\mu = 0\farcs0056\pm 0\farcs0009\, {\rm d}^{-1} = 2\farcs0\pm 0\farcs3\, {\rm yr}^{-1}$.
This proper motion corresponds to the projected velocity $v_\perp =(0.07\pm 0.01)c$ at $d=2.3$ kpc.
The launch time, defined as $r(t_{\rm launch}) =0$, is $t_{\rm launch} = -21^{+122}_{-165}$ days. This range of $t_{\rm launch}$ 
encompasses the periastron 
and the {\sl Fermi} flare dates.

\subsection{Spectral analysis}

We extracted the spectra from each of the two new observations and 
fitted them simultaneously.
The extraction regions are shown in Figure \ref{pic8}. We fit the spectra of the compact core 
(6257 counts
in the 0\farcs9 radius aperture)
and the blob (58 counts in the 3.6 arcsec$^2$ area) with the
absorbed power-law model, using the XSPEC {\tt phabs} model for absorption.
We obtained the following fitting parameters for the core:
photon index $\Gamma=1.5\pm 0.4$,
hydrogen column density $N_H=(4.2\pm 0.3)\times 10^{21}$ cm$^{-2}$,
and unabsorbed flux $F_{\rm 0.5-8\,keV}^{\rm unabs}=(1.95\pm 0.04)\times 10^{-12}$ erg cm$^{-2}$ s$^{-1}$ { (see Table 1 for more details)}.

\begin{figure}
\centering
\includegraphics[width=9cm]{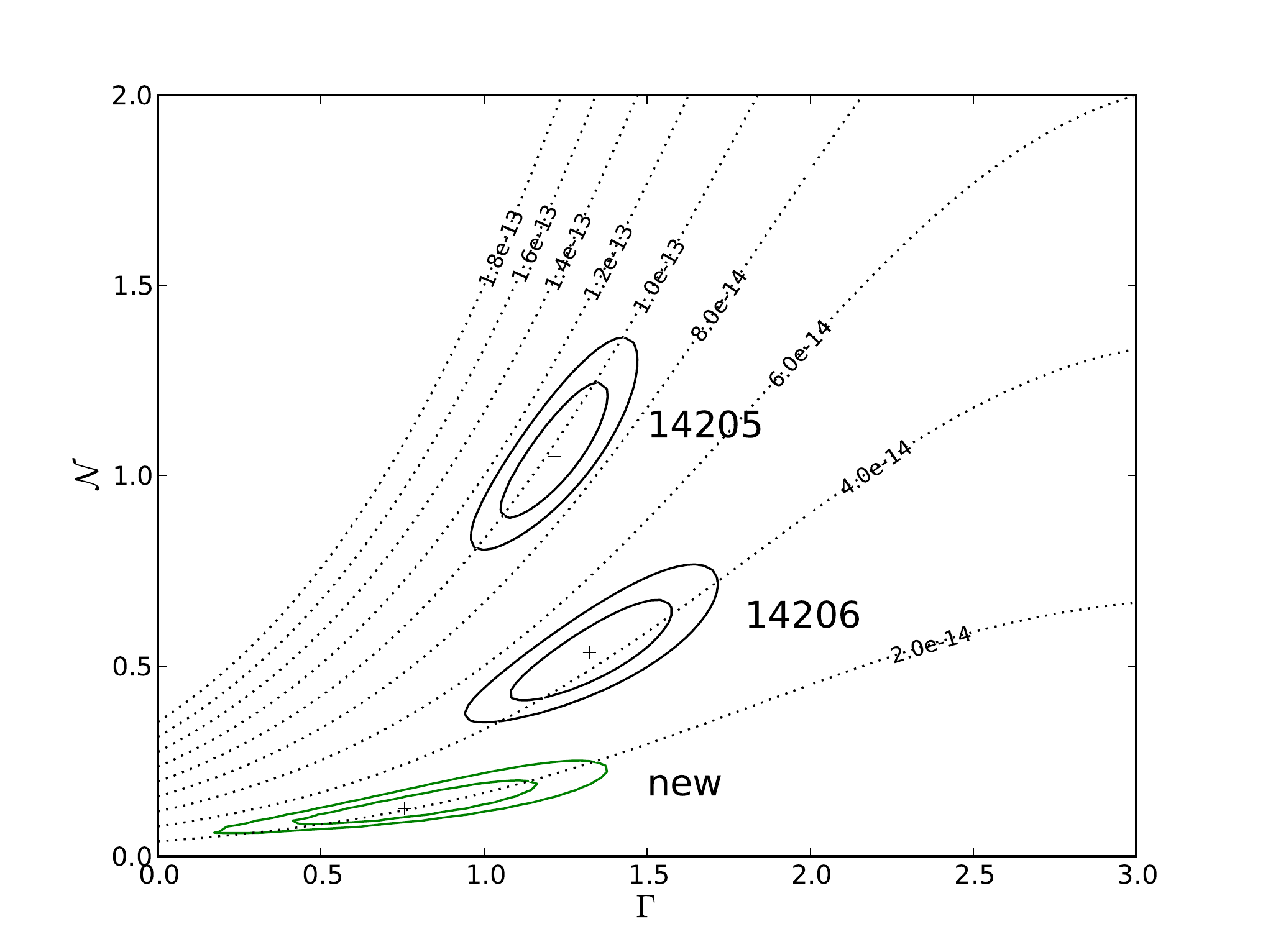}
\caption{
Confidence contours (90\%, and 99\%) in the $\Gamma$--${\cal N}$ plane for the ACIS spectra of the extended nebula (see Figure \ref{pic2}
and K+14 for region
definitions) computed for the absorbed PL model. The PL normalization ${\cal N}$ is in units of $10^{-5}$
photons cm$^{-2}$ s$^{-1}$ keV$^{-1}$ at 1 keV.
 The dashed lines are lines of constant unabsorbed flux in
the 0.5--8 keV band (labeled in erg\,cm$^{-2}$\,s$^{-1}$). The contours for observations 14205, 14206 and the new data were fit with $N_H$ fixed at the values of $N_{\rm H} = 2.9$, 3.1, 3.0 $\times$ 10$^{21}$ cm$^{-2}$, respectively.
\label{pic6}
}
\end{figure}

Because of the small number of counts in the blob,
we used C-statistic\footnote{https://heasarc.gsfc.nasa.gov/xanadu/xspec/manual/
XSappendixStatistics.html} 
(without binning) and fixed $N_{\rm H}$ to fit its spectrum. 
{ The fitting parameters are provided in Table 1.}
Figure \ref{pic6} shows the 90\% and 99\% confidence contours in the 
$\Gamma$-${\cal N}$ plane (${\cal N}$ is the power-law normalization)
 for $N_H=3\times 10^{21}$ cm$^{-2}$.
The latter value is smaller than the one we obtained from fitting the
core spectrum; we chose it as the average of the $N_{\rm H}$ values obtained in the fits of the core spectrum in ObsIDs 14205 and 14206 (K+14) in order to exclude additional absorption from within the binary itself.
Figure \ref{pic6} also shows the $\Gamma$-${\cal N}$ confidence contours for 
the extended emission from two previous observations, which we re-fit
using the C-statistic for consistency.
We see that the photon index for the blob in the new data, $\Gamma = 
0.8 \pm 0.4$, 
does not exceed those found in the previous observations,
$\Gamma = 1.2\pm0.2$ and $1.3\pm 0.2$ for ObsIDs 14205 and 14206,
respectively (i.e., 
the spectral evolution does not show any cooling of the blob's matter).
We see from Figure \ref{pic6} that the flux of extended emission has been
steadily decreasing. Fitting it with
an exponential decay law 
yields a characteristic decay time of 
$540 \pm  100$ days.

\section{Discussion}

K+14 have discussed  possible scenarios  regarding the origin and nature 
of the observed variable extended emission. One possibility considered by K+14 
was the synchrotron emission from the PW that escapes from the binary in the direction away from the
high-mass star and shocks in the extrabinary medium,
similar to 
PWNe of young solitary pulsars 
\citep{2008AIPC..983..171K}.
The estimated stand-off distance to such a shock 
is comparable to the observed separations for 
{ an ambient pressure $p_{\rm amb}\sim 10^{-10}$ dyn cm$^{-2}$}.
Because of the large eccentricity of the binary, the pulsar spends most of orbital cycle around 
apastron, so most of the shocked PW matter
concentrates around the extension of the binary's major axis in the apastron
direction.
The separation between the extrabinary shock and the pulsar could vary due to changes of the
ratio of the PW pressure
to the ambient pressure.
However, this interpretation looks somewhat artificial now because
the new data strongly suggest that the separation is steadily increasing, 
which would require a steadily decreasing ambient pressure (or increasing
PW pressure) on a time scale of $\sim 800$ days.
In addition, { the required ambient pressure is unrealistically high
for a circumbinary medium, which is likely filled with a freely flowing
stellar wind of the massive companion, and} the extended structure does not have an arc-like appearance 
anymore, hence losing its resemblance to a termination shock.

 Another scenario considered by K+14 assumed that the moving structure is a clump of matter
ejected from the binary by the pulsar interaction with the decretion disk.
The new data seem to support this scenario because the average projected velocity implies that the structure could indeed be launched near periastron.
   However, as K+14  pointed out, the fast, steady motion over such a large period of time poses a number of problems.
If the ejected clump consists solely of synchrotron-emitting  
relativistic electrons or positrons, possibly confined by the magnetic field,
it should decelerate very quickly by the drag force (assuming it moves through a much slower expanding medium, such as a bubble created by the companion's wind).  
On the other hand, if the clump consists solely of electron-ion plasma
(e.g., a fraction of the decretion disk heated and expelled from the binary
by the interaction with the PW)\footnote{In this case, the clump emission
would be
thermal bremsstrahlung.
As shown by K+14, the
spectra of the extended feature can be fitted by a thermal bremsstrahlung 
model with $kT\gtrsim 10$ keV, $n_e\sim n_p\sim 10^2$ cm$^{-3}$.}, then
its mass would be $m_{\rm cl}\sim 10^{29}$ g,
 much larger than the plausible mass of the disk,
$m_{\rm disk}\sim 10^{24}-10^{26}$ g (\citeauthor{2014MNRAS.439..432C} \citeyear{2014MNRAS.439..432C}, and references therein),
or  the mass $\dot{M}P_{\rm orb}=6.7\times 10^{25} (\dot{M}/10^{-8}\,M_\odot/{\rm yr}$) g
lost by the high-mass companion
during one orbital cycle.
In addition, its kinetic energy, $E_{\rm cl}\sim 10^{47}$ erg,
would exceed the energy that could be supplied by any conceivable source.

A mixed scenario, in which the emission is produced by relativistic electrons 
supplied by the PW but the clump's mass is determined by protons/ions from the stellar
outflow,
  may be more plausible. 
However, even in this scenario, { the clump's mass should be a 
substantial fraction of $m_{\rm disk}$ 
to overcome the drag force, even at extremely low ambient densities
(K+14).}
Moreover, it remains unclear what is the source of the corresponding
clump's kinetic energy, $E_{\rm cl}
\sim 5\times 10^{43} (m_{\rm cl}/10^{25}\,{\rm g})(v/0.1c)^2$ erg,
let alone the acceleration mechanism.
If the
clump's energy is provided by the loss of a small fraction of the
pulsar's spin energy, $E_{\rm sp} = 8.6\times 10^{48}$ erg
(for the moment of inertia $I=10^{45}$ g cm$^2$),
during the disk
passage,
the decrease of the spin
energy, $\Delta E_{\rm sp} \sim E_{\rm cl}$, would lead to a perceptible increase
of the spin period, $\Delta P \sim (P^3/4\pi^2 I) E_{\rm cl} \sim 
6\times 10^{-8}
(m_{\rm cl}/10^{25}\,{\rm g})$ s.
However, no such period jumps were noticed in the most detailed timing
analysis available
 \citep{2014MNRAS.437.3255S}.
These authors mention
systematic variations in timing residuals close to the 6 periastron passages
covered by their analysis, but they attribute those to uncorrected
variations in dispersion measure or scattering of the pulsar radiation through
the disk.
Thus, it seems
that the spin-down
due to the pulsar-disk interaction does not look as a plausible source
of  the clump energy.
In principle, one might assume that a small fraction of pulsar's kinetic energy,
$E_{\rm kin}\sim 10^{48}$ erg (near periastron), is somehow converted
into the clump's energy in the course of pulsar-disk interaction, but a decrease of
$E_{\rm kin}$ should result in a decrease of binary period $P_b$, which is not
consistent with the results of timing analysis\footnote{
On the contrary, \cite{2014MNRAS.437.3255S} found a hint of {\em increase} of the
binary period, with a derivative $\dot{P}_b = (1.4\pm 0.7)\times 10^{-8}$ s s$^{-1}$, corresponding to $\Delta P_b \approx 1.5$ s per binary cycle,
which they attribute to mass loss due to the polar wind of LS\,2883.
}.
Thus, we conclude that even the mixed clump scenario looks problematic
because
ejection of a large mass is required to  overcome
the drag force, and it remains unclear what is the energy source
and how such a mass could be accelerated to a velocity $\sim 0.1 c$
during the short time of the pulsar-disk interaction.

The drag deceleration and energy deficit problems can be alleviated 
if, instead of the companion's wind bubble,
 {\em the clump is moving 
in the unshocked PW}
whose velocity, close to the speed of light,
significantly exceeds the clump's velocity. 
This scenario 
becomes more { plausible}
at larger 
values of
the parameter $\eta=\dot{E}/(\dot{M}v_{w}c) =4.4\, \dot{M}_{-9}^{-1} v_{w,8}^{-1}$,
where $\dot{M}_{-9}$ is the mass-loss
rate in the O-star polar wind\footnote{{ The reference value of mass-loss 
rate, 
$\dot{M}=10^{-9}\,\, M_\sun\, {\rm yr}^{-1}$, is lower than usually adopted
for LS\,2883, but such a low $\dot{M}$ is consistent with observations of
OV stars with luminosities $L_*<10^{5.3} L_\sun$,
which show mass losses lower than theoretically predicted or expected from
extrapolations from more luminous O stars (the so-called ``weak wind problem'';
see Section 5 in the review by \citeauthor{2008A&ARv..16..209P} \citeyear{2008A&ARv..16..209P}.) }}
  in units of $10^{-9}\,\, M_\sun\, {\rm yr}^{-1}$,
and $v_{w,8}$ is the wind's velocity in units of 1000 km s$^{-1}$.
At $\eta >1$, the companion's wind 
is confined by the PW into
a cone with half-opening angle
$\alpha\approx 30^\circ (4-\eta^{-2/5})\eta^{-1/3}$ \citep{1993ApJ...402..271E},
while the unshocked PW fills
the rest of circumbinary volume\footnote{{ A large value of $\eta$ is needed in our scenario because in the opposite case
the unshocked PW would occupy only a narrow
channel confined by the stellar wind and rotating around the massive star
together with the pulsar.
Even if a clump of
stellar matter gets into that channel, it will go out from the channel in
a fraction of the binary period and will be decelerated by the stellar
wind.} 
We note that larger $\eta$ also imply the intra-binary
shock is closer to the massive star,
which helps to explain the high-energy emission by the
Compton up-scattering of stellar
photons \citep{2008MNRAS.385.2279S}.}.
{ Since the pressure in the freely flowing (unshocked) polar wind of the high-mass
companion is proportional to $r^{-2}$, similar to the PW ram pressure,
$p_{\rm pw} = \dot{E}/(4\pi c r^2) = 2.2\times 10^{-10} r_{17}^{-2}$ dyn cm$^{-2}$ (where 
$r_{17} = r/10^{17}$ cm), the PW cannot shock until it reaches the
termination shock of the stellar wind at $r=r_{\rm ts}$, where the 
wind bubble pressure jumps up by a few orders of magnitude
(see Figure 1 in \citeauthor{2005ApJ...630..892D} \citeyear{2005ApJ...630..892D}).
This means that the PW} 
 shocks at large distances from the binary
{ (e.g, at $\sim 3'$ for $r_{\rm ts}\sim 2$ pc)},
where the surface brightness
of the shocked PW emission (PWN) may be too low to be detected.
In this scenario,
the observed extended X-ray emission
can be due to synchrotron radiation of the PW shocked by
the collision
with the clump of ejected material. The X-ray luminosity can be estimated
as $L_{\rm X,cl}=\eta_X \dot{E} (r_{\rm cl}/2 r)^2$,
where $r$ is the distance from the pulsar, $r_{\rm cl}$ is the effective
radius of the clump, and $\eta_X$ is the X-ray efficiency. For instance,
$L_{\rm X,cl}\sim 
0.01\eta_X\dot{E}\sim 
8\times 10^{33} \eta_X$ erg s$^{-1}$ for $r_{\rm cl}/r \sim 
0.2$
for the latest observation.
This estimate
is consistent with the observed X-ray luminosity,
$L_{\rm X,cl}=(1.3\pm 0.3)\times 10^{31}d_{2.3}^2$ erg s$^{-1}$,
 at
a reasonable value for X-ray efficiency, $\eta_X\sim 
1.5\times 10^{-3}$.
This interpretation of the X-ray emission
 is also consistent with the lack of spectral
softening 
(Section 3.2) because
cooling of relativistic electrons is compensated by the energy
supplied by the PW.

\begin{deluxetable*}{ccccccccccccc}
\tabletypesize{\footnotesize} 
\tablecolumns{13} 
\tablewidth{0pt} 
\tablewidth{0.99\textwidth}
\tablecaption{Spectral fit parameters for the core and extended emission in three \cxo ACIS-I  observations} 
\tablehead{
\colhead{ObsID} & \colhead{MJD} & \colhead{$\theta$\tablenotemark{a}} & \colhead{$\Delta t$\tablenotemark{b}} & \colhead{Exp.\tablenotemark{c}} & \colhead{Cts\tablenotemark{d}} & 
\colhead{$F_{\rm obs}$\tablenotemark{e}} & \colhead{$F_{\rm corr}$\tablenotemark{f}} & \colhead{$N_H$} & \colhead{$\Gamma$} & \colhead{$\mathcal{N}$\tablenotemark{g}} &
\colhead{$\mathcal{A}$\tablenotemark{h}} & \colhead{$\chi^2/$dof} \\ 
\colhead{} & \colhead{}  & \colhead{deg} & \colhead{days} & \colhead{ks} & \colhead{}  & \colhead{$10^{-14}$ cgs} & \colhead{$10^{-14}$ cgs} & \colhead{$10^{21}$ cm$^{-2}$}  &\colhead{}  &\colhead{$10^{-4}$} &\colhead{ (arcsec$^2$)} & \colhead{}}
\startdata 
10089 & 54965& 182 &667 &25.6 & 1825 & 139(5) & 158(6) & 1.5(7) & 1.51(10) & 2.3(3) & 2.5 & 20.9/30\\
  & & & & & 61 & 2.8(1) & 3.1(1) & 1.5$^\ast$ & 1.3(5) & 0.039(16) & 22.1 & 3.1/9 \\
14205& 55912 & 169 &370 & 56.3& 6551 & 249(4) & 296(5) & 2.9(3) & 1.39(5) & 3.8(2) & 2.5 & 76.6/87\\
    & & & & & 343 & 8.5(5) & 9.2(7) & 2.9$^\ast$ & 1.2(1)& 0.10(1) & 22.1 & 36.24\%\tablenotemark{k}\\
14206& 56431 & 192 &  886 & 56.3 &  4162 & 137(5) & 176(7) & 3.1(3) & 1.68(6) & 3.07(2) & 2.5 & 146/169 \\
   & & & & &  144 &  3.6(4) & 3.9(6) & 3.1$^\ast$ &  1.3(2) &  0.052(14) & 12.8  & 62.96\%\tablenotemark{k}\\
$New$\tablenotemark{i} & 56696 & 221 & 1151 & 57.6 & 6257 &  149(5) & 195(4) & 4.2(3) & 1.5(4) & 3.0(1) & 2.5 & 242.5/204 \\
   & & & & & 58 & 
1.9(4)  & 2.0(4) & 3.0*\tablenotemark{j}  & 0.8(4) & 0.013(7) & 3.6 & 54.12\%\tablenotemark{k}   
\enddata 
\tablecomments{ For each ObsID  the upper and the lower rows correspond to  the partially resolved core and the extended emission, respectively.The $1\sigma$ uncertainties are shown in parentheses  (see also Figure 4). The XSPEC extinction model {\tt phabs} was used throughout. Fluxes and counts are in the 0.5--8\,keV range, corrected for finite aperture size for the core emission. An asterisk indicates that the extinction column was fixed to the value of the corresponding point source fit.}
\tablenotetext{a}{True anomaly counted from periastron.}
\tablenotetext{b}{Days since  latest preceded periastron.}
\tablenotetext{c}{Exposure corrected for deadtime.}
\tablenotetext{d}{Total (gross) counts.}
\tablenotetext{e}{Observed flux.}
\tablenotetext{f}{Extinction corrected flux.}
\tablenotetext{g}{Normalization in photons\,s$^{-1}$\,cm$^{-2}$\,keV$^{-1}$ at 1\,keV.}
\tablenotetext{h}{Area of the extraction region.}
\tablenotetext{i}{ObsIds 16563 and 16583 fit simultaneously.}
\tablenotetext{j}{Fit was done, for extended emission, using the fixed $N_{\rm H}$ value obtained from fitting the core around apastron passage to exclude any absorption from within the binary.}
\tablenotetext{k}{C-statistics were used for fitting the extended
emisssion spectra across all observations for consistency.
This percentage is the Null Hypothesis probability 
$1-{\cal P}$,  where ${\cal P}$ is the probability given by
the XSPEC {\tt goodness} command with 10,000 runs.} 
\end{deluxetable*}

The interaction of
the unshocked PW with the ejected fragment is not only responsible for
its X-ray emission, but it can also {\em accelerate} the fragment.
A very crude
estimate of acceleration
 is $\dot{v} \sim p_{\rm pw} A m_{\rm cl}^{-1} \sim 
440 r_{17}^{-2}\xi_A (m_{\rm cl}/10^{21}\,{\rm g})^{-1}$ cm s$^{-2}$,
where $p_{\rm pw}$
is the PW ram pressure,
$A \sim 2\times 10^{33} \xi_A$ cm$^2$ is the clump's
cross section, and $\xi_A<1$ is the filling factor\footnote{The `clump' can consist of a few smaller fragments, which we cannot resolve in the image. These fragments could
be launched from the binary in slightly different directions,
 which could
explain the large size of the observed extended feature. }.
Interestingly, this
estimate is consistent with the apparent acceleration estimated
(with low significance) in
Section 3.1 at $m_{\rm cl} \sim  10^{21} 
\xi_A$ g (for $r=2\times 10^{17}$ cm, i.e., $\sim 7''$ from the pulsar).
Ejection of such a small mass should not make a measurable effect
on pulsar timing.
Moreover, the possibility of gradual acceleration of the
ejected clump alleviates the problem of a huge acceleration of the
clump in the course of its ejection caused by pulsar-disk
interaction. For instance, for $m_{\rm cl}\sim 10^{21}$ g, the clump's
kinetic energy, $E_{\rm cl}\sim 4.5\times 10^{39} (v/0.1c)^2$ erg,
is a tiny fraction, $\sim 5\times 10^{-5}$,
 of the pulsar's spin-down energy loss during, e.g., the time elapsed
from the 2010 periastron.
We should note that if the acceleration during
our observations was indeed so large, then the conclusion
that the clump was launched around the 2010 periastron, derived under
the assumption of constant speed,
becomes questionable (unless the
acceleration was much lower
when the clump was closer to the binary,
e.g., due to a larger accelerated mass).
The matter could be ejected by the pulsar-disk interaction
close to an earlier periastron, but it is hardly possible to estimate
the launch time without detailed modeling of the PW-clump
interaction, accounting for varying acceleration.

A qualitative scenario for cloud formation
and acceleration might look as follows.
Before entering the disk, 
the pulsar moves in the fast, radially directed companion's polar
wind, the apex of the intrabinary bow shock is on the line connecting the pulsar
and the companion, and the shock position and shape are determined by the parameter
$\eta$
\citep{2013A&ARv..21...64D}.
When the pulsar enters the disk, 
the structures of both the disk and
 the intrabinary
shock change dramatically.
 Since the matter in the disk
moves in almost circular orbits around the companion with nearly Keplerian
velocity \citep{2013A&ARv..21...69R}, which is comparable to the orbital
velocity of the pulsar,
 the relative velocity of the pulsar and the
disk matter, ${\bf v}_{\rm rel} = {\bf v}_{\rm disk} - {\bf v}_{\rm psr}$,
 is not directed radially, and
 the shock apex is not on the pulsar-companion line
(see an illustration in \citeauthor{2012ApJ...752L..17K} \citeyear{2012ApJ...752L..17K}).
The (minimum) 
distance of the shock from the pulsar can be estimated 
as
$r_s\sim \dot{E}^{1/2}[4\pi c\rho_{\rm disk}(R)]^{-1/2} |{\bf v}_{\rm rel}(R)|^{-1}$,
where $\rho_{\rm disk}(R)
\sim \rho_{0}(R_*/R)^3$ 
is the density  
in the mid-plane of the disk in the pulsar vicinity (at the distance $R$ from
the companion),
$R_*\sim 10R_\sun$ is the companion's radius. 
Using the same disk model as \cite{2014MNRAS.439..432C}, we obtain the shock radius
$r_s\sim 1.5\times 10^{13} (\rho_{\rm disk}/10^{-16}\,{\rm g\,cm}^{-3})^{-1/2} (v_{\rm rel}/100\,{\rm km\,s^{-1}})^{-1}$ cm, for the first crossing of the disk  ($R\sim 40 R_*\sim 2$\,AU).
Since $r_s$ exceeds
(or is comparable to) the vertical size of the disk, $H(R=40 R_*) \sim 5\times 10^{12}$ cm, 
the disk should be 
significantly disrupted by the first pulsar passage, 
which is consistent with numerical simulations
\citep{2011PASJ...63..893O, 2012ApJ...750...70T}. 
In the second disk passage further fragmentation of the disk  can occur.
The collision of the PW with disk matter fragments would result in shocks and
$\gamma$-ray flares from the shocked PW, as well as  entrainment of the
clumps in the PW flow, which would blow them out of the binary
and eventually accelerate up to the observed velocity $\sim 0.1 c$. 
The details of the mass loading and ejection processes are likely
very complex and should be investigated with numerical simulations. 

A possible mechanism of matter ejection from the binary
could be connected with episodic
accretion of the disk matter onto the neutron star
magnetosphere\footnote{Accretion onto the neutron star surface would be inconsistent with the relatively low X-ray luminosity of B1259 even around periastron.}. 
If there were no PW and associated shock, the neutron star moving in the decretion disk could
gravitationally capture
the matter within the cylinder of radius $R_{\rm grav}=
2GM_{\rm psr} v_{\rm rel}^{-2} = 3.7\times 10^{12} (v_{\rm rel}/100\,{\rm km\,s}^{-1})^{-2}$ cm (for $M_{\rm psr}=1.4 M_\sun$), with the
 maximum possible capture rate $\dot{m}_{\rm max}=\pi R_{\rm grav}^2 \rho_{\rm disk}v_{\rm rel} = 4.4\times 10^{16} (\rho_{\rm disk}/10^{-16}\,{\rm g\,cm}^{-3}) (v_{\rm rel}/100\,{\rm km\,s}^{-1})^{-3}$ g s$^{-1}$. However, if the radius $r_s$ of the
PW shock exceeds $R_{\rm grav}$ (as for the parameters assumed
above for the shock radius estimate), the matter is deflected by the
shock out of the capture cylinder.
Nevertheles, the capture becomes possible if $R_{\rm grav}>r_s$, i.e.,
if $\rho_{\rm disk}>1.6\times 10^{-15} (v_{\rm rel}/100\,{\rm km\,s}^{-1})^2$ g cm$^{-3}$. 
If the disk density is indeed so high 
in the vicinity of the crossing pulsar (which is not ruled out --- see
\citeauthor{2012ApJ...750...70T} \citeyear{2012ApJ...750...70T}), then the
ram pressure of the gravitationally attracted
 matter ($\rho v_{\rm ff}^2\propto r^{-5/2}$) overcomes the ram
 pressure of the PW ($p_{\rm pw}\propto r^{-2}$), and the disk matter streams toward the
pulsar with a nearly free-fall velocity, $v_{\rm ff}=(2GM_{\rm psr}/r)^{1/2}$,
 until it 
reaches the pulsar magnetosphere at $r\sim
R_{\rm lc}=cP/(2\pi) = 2.3\times 10^8$ cm. 
At radii smaller than the light cylinder radius $R_{\rm lc}$ the accreting
matter is stopped by
the magnetic
field pressure ($B^2/8\pi \propto r^{-6}$) in the pulsar magnetosphere, at 
the Alfven radius $R_A\sim (8GM_{\rm psr})^{-1/7} \mu^{4/7} \dot{m}^{-2/7}
\sim 1.4\times 10^8\dot{m}_{17}^{-2/7}$ cm, where $\mu=B_{\rm NS}R_{\rm NS}^3
\sim 3.3\times 10^{29}$ G cm$^3$ is the pulsar's magnetic moment, and 
$\dot{m}_{17} = \dot{m}/(10^{17}\,{\rm g\,s}^{-1})$. As long as $R_A$ exceeds
the corotation radius $R_{\rm cor}= (GMP^2/4\pi)^{1/3} = 2.2\times 10^7$ cm
(i.e., at any reasonable accretion rate $\dot{m}<6\times 10^{19}$ g s$^{-1}$),
accretion occurs in the `propeller regime' \citep{1975A&A....39..185I}, i.e.,
the accreting matter does not reach the neutron star surface.
Most of this matter is expelled in a disk-shaped outflow in the equatorial plane
 by the centrifugal forces (see \citeauthor{2003ApJ...588..400R} 
\citeyear{2003ApJ...588..400R} for MHD simulations) 
while up to $\sim 10\%$ 
can leave the system in a high-velocity jet
 along the neutron star spin
axis\footnote{In the numerical similations of matter flows in the propeller
regime (e.g., \citeauthor{2009MNRAS.399.1802R} \citeyear{2009MNRAS.399.1802R}) 
only the case of aligned magnetic and spin
axes was considered. We presume that the jet formation can also
occur if the axes are misaligned, when the jet should be directed along the
the spin axis by the symmetry requirements.}
 (\citeauthor{2014CAC.....1....3L} \citeyear{2014CAC.....1....3L}, 
and references therein), with a velocity up to $(GM/R_A)^{1/2} \sim 
0.04 \dot{m}_{17}\,c$.
An upper limit on the total mass ejected
in a jet during the disk passage can be as large as $\sim 0.1\dot{m}t_{\rm disk} \sim 10^{22} \dot{m}_{17}(t_{\rm disk}/15\,{\rm d})$ g.
If the ejected matter gets into the region of unshocked PW, it can be
further accelerated to the observed velocity of $\sim 0.1c$.
  Thus, 
accretion of decretion disk matter in the propeller regime 
could, in principle, be responsible for ejection of the observed clump,
but it would
require a rather high accretion rate (i.e., a very large disk density).

Overall, the interpretation of the moving extended X-ray feature 
as a fragment of the decretion disk ejected from the binary by the
pulsar-disk interaction and moving in the unshocked PW looks quite 
plausible. However, we should mention a possible contradiction
between such an interpretation and the
detections of extended radio emission around the pulsar
29 days after the 2010 periastron 
\citep{2014MNRAS.439..432C} and 21 days after the 2007 periastron
\citep{2011ApJ...732L..10M}.
That emission, with a total size of
$\sim 50$ mas
(115 AU at $d=2.3$ kpc)
and a larger extension toward northwest, 
 was interpreted as a cometary tail extending behind the
pulsar in the northwest direction,
quite different from
the southwest direction of the X-ray feature motion.
To reconcile the different directions, we have to assume that the
radio and X-ray features are intrinsically different (e.g., comprised
of different matter) or that the ejected fragment could change the
direction of its motion because of the varying direction of the accelerating
force close to the binary.

 \section{Conclusions}
Our new {\sl CXO} observation of the 
extended object
in the vicinity of the B1259
binary has shown that 
the 
object keeps moving away from the binary.
The comparison with  two previous observations shows that its
luminosity has been steadily decreasing, and its
average projected velocity was $v_\perp\approx 0.07c$ during
780 days covered by the observations.
If that velocity has remained constant in the course of motion, the 
object was ejected from the binary near the 2010 periastron, 
perhaps by the collision of the pulsar with the decretion disk of the
high-mass companion.
If this clump of matter were moving in a slowly expanding circumbinary medium, 
its mass and kinetic energy would be uncomfortably
 large to overcome the drag deceleration.
Therefore, we suggest the clump (likely a fragment of the decretion disk)
is moving in the unshocked relativistic PW. Its X-ray emission can be
interpreted as synchrotron radiation of the PW shocked by an interaction 
with the fragment. The ram pressure of the PW can be responsible for
clump acceleration, a hint of which is seen in our observations. 
No such phenomena have been observed so far.
Further monitoring with {\sl CXO} is needed to confirm or refute this
interpretation.

\acknowledgements
Support for this work was provided by the National Aeronautics and Space Administration through {\sl Chandra} Awards DD3-14070 and GO2-13085 
issued by the {\sl Chandra} X-ray Observatory Center, which is operated by the Smithsonian Astrophysical Observatory for and on behalf of the National Aeronautics Space Administration under contract NAS8-03060. We are very grateful to Harvey Tananbaum  for allocating the DDT time to continue monitoring this system.
We also thank Dmitry Khangulyan for useful discussions { and the anonymous
referee for careful reading of the paper and useful comments}.

{\it Facility:} \facility{CXO}


\begin{thebibliography}{}


\bibitem[Abdo et al.(2011)]{2011ApJ...736L..11A} Abdo, A.~A., Ackermann, 
M., Ajello, M., et al.\ 2011, \apjl, 736, L11

\bibitem[Aharonian et 
al.(2005)]{2005A&A...442....1A} Aharonian, F., Akhperjanian, A.~G., Aye, K.-M., et al.\ 2005, \aap, 442, 1

\bibitem[Aharonian et 
al.(2009)]{2009A&A...507..389A} Aharonian, F., Akhperjanian, A.~G., Anton, G., et al.\ 2009, \aap, 507, 389

\bibitem[Ball et al.(1999)]{1999ApJ...514L...39B} Ball, l., Melatos, A., Johnston, S., \& Skj{\ae} Raasen, O. 1999, ApJL, 514, L39


\bibitem[\protect\citeauthoryear{{Chernyakova} et~al.}{{Chernyakova}
  et~al.}{2006}]{2006MNRAS.367.1201C}
{Chernyakova}, M., {Neronov}, A., {Lutovinov}, A., {Rodriguez}, J.,  \&
  {Johnston}, S. 2006, MNRAS, 367, 1201
  
 \bibitem[Chernyakova et al.(2009)]{2009MNRAS.397.2123C} Chernyakova, M., 
Neronov, A., Aharonian, F., Uchiyama, Y., 
\& Takahashi, T.\ 2009, \mnras, 397, 2123


\bibitem[Chernyakova et al.(2014)]{2014MNRAS.439..432C} Chernyakova, M., 
Abdo, A.~A., Neronov, A., et al.\ 2014, \mnras, 439, 432


\bibitem[Dubus(2013)]{2013A&ARv..21...64D} Dubus, G.\ 2013, \aapr, 21, 64


\bibitem[Dwarkadas(2005)]{2005ApJ...630..892D} Dwarkadas, V.~V.\ 2005, 
\apj, 630, 892 

\bibitem[Eichler \& Usov(1993)]{1993ApJ...402..271E} Eichler, D., \& Usov, V. 1993, \apj, 402, 271


\bibitem[Illarionov 
\& Sunyaev(1975)]{1975A&A....39..185I} Illarionov, A.~F., \& Sunyaev, R.~A.\ 1975, \aap, 39, 185 

\bibitem[\protect\citeauthoryear{{Johnston} et~al.}{{Johnston}
  et~al.}{1992}]{1992ApJ...387L..37J}
{Johnston}, S., {Manchester}, R.~N., {Lyne}, A.~G., {Bailes}, M., {Kaspi},
  V.~M., {Qiao}, G.,  \& {D'Amico}, N. 1992, \apjl, 387, L37


\bibitem[\protect\citeauthoryear{{Johnston} et~al.}{{Johnston}
  et~al.}{2005}]{2005MNRAS.358.1069J}
{Johnston}, S., {Ball}, L., {Wang}, N.,  \& {Manchester}, R.~N. 2005, \mnras,
  358, 1069

\bibitem[\protect\citeauthoryear{{Kargaltsev} \& {Pavlov}}{{Kargaltsev} \&
  {Pavlov}}{2008}]{2008AIPC..983..171K}
{Kargaltsev}, O.,  \& {Pavlov}, G.~G. 2008, in American Institute of Physics
  Conference Proceedings, Vol. 983, 40 Years of Pulsars: Millisecond Pulsars,
  Magnetars and More, ed. C.~{Bassa}, Z.~{Wang}, A.~{Cumming}, \& V.~M.
  {Kaspi}, 171

\bibitem[Kargaltsev et al.(2014)]{2014ApJ...784..124K} Kargaltsev, O., 
Pavlov, G.~G., Durant, M., Volkov, I., \& Hare, J.\ 2014, \apj, 784, 124 (K+14)

\bibitem[Khangulyan et al.(2007)]{2007MNRAS.380..320K} Khangulyan, D., 
Hnatic, S., Aharonian, F., \& Bogovalov, S.\ 2007, \mnras, 380, 320 

\bibitem[Khangulyan et al.(2012)]{2012ApJ...752L..17K} Khangulyan, D., Aharonian, F.\ A., Bogovalov, S.\ V., Rib\'{o}, M. 2012, \apj, 752, L17


\bibitem[Lovelace et al.(2014)]{2014CAC.....1....3L} Lovelace, R.~V., 
Romanova, M.~M., Lii, P., 
\& Dyda, S.\ 2014, Computational Astrophysics and Cosmology, 1, 3 

\bibitem[Malyshev et al.(2014)]{2014ATel.6204....1M} Malyshev, D., Neronov, 
A., \& Chernyakova, M.\ 2014, The Astronomer's Telegram, 6204, 1 



\bibitem[\protect\citeauthoryear{{Melatos}, {Johnston}, \& {Melrose}}{{Melatos}
  et~al.}{1995}]{1995MNRAS.275..381M}
{Melatos}, A., {Johnston}, S.,  \& {Melrose}, D.~B. 1995, \mnras, 275, 381

\bibitem[Mold\'{o}n et al.(2011)]{2011ApJ...732L..10M} 
Mold\'{o}n, J., Johnston, S., Rib\'{o}, M., Paredes, J.\ M., \& Deller, A.\ T. 2011, \apj, 732, L10

\bibitem[\protect\citeauthoryear{{Negueruela} et~al.}{{Negueruela}
  et~al.}{2011}]{2011ApJ...732L..11N}
{Negueruela}, I., {Rib{\'o}}, M., {Herrero}, A., {Lorenzo}, J., {Khangulyan},
  D.,  \& {Aharonian}, F.~A. 2011, \apjl, 732, L11

\bibitem[Okazaki et al.(2011)]{2011PASJ...63..893O} Okazaki, A.~T., 
Nagataki, S., Naito, T., et al.\ 2011, \pasj, 63, 893 

\bibitem[\protect\citeauthoryear{{Pavlov}, {Chang}, \& {Kargaltsev}}{{Pavlov}
  et~al.}{2011}]{2011ApJ...730....2P}
{Pavlov}, G.~G., {Chang}, C.,  \& {Kargaltsev}, O. 2011, ApJ, 730, 2

\bibitem[Puls et al.(2008)]{2008A&ARv..16..209P}
Puls, J., Vink, J.\ S., Najarro, F. 2008, A\&A Rev., 16, 209

\bibitem[Rivinius et 
al.(2013)]{2013A&ARv..21...69R} Rivinius, T., Carciofi, A.~C., \& Martayan, C.\ 2013, \aapr, 21, 69 

\bibitem[Romanova et al.(2003)]{2003ApJ...588..400R}
Romanova, M.\ M., Toropina, O.\ D., Toropin, Yu.\ M., Lovelace, R.\ V.\ E.
2003, \apj, 588, 400

\bibitem[Romanova et al.(2009)]{2009MNRAS.399.1802R}  Romanova, M.\ M., Ustyugova, G.\ V., Koldoba, A.\ V., \& Lovelace, R.\ V.\ E. 2009, \mnras, 399, 1802

\bibitem[Shannon et al.(2014)]{2014MNRAS.437.3255S} Shannon, R.~M., 
Johnston, S., \& Manchester, R.~N.\ 2014, \mnras, 437, 3255 

\bibitem[Sierpowska-Bartosik \& Bednarek(2008)]{2008MNRAS.385.2279S} Sierpowska-Bartosik, A., \& Bednarek, W. 2008, \mnras, 385, 2279

\bibitem[Takata et al.(2012)]{2012ApJ...750...70T} Takata, J., Okazaki, 
A.~T., Nagataki, S., et al.\ 2012, \apj, 750, 70 

\bibitem[Tam et al.(2011)]{2011ApJ...736L..10T} Tam, P.~H.~T., Huang, 
R.~H.~H., Takata, J., et al.\ 2011, \apjl, 736, LL10 

\bibitem[Tam et al.(2015)]{2015ApJ...798L..26T} Tam, P.~H.~T., Li, K.~L., 
Takata, J., et al. 2015, ApJL, 798, L26

\bibitem[Tavani 
\& Arons(1997)]{1997ApJ...477..439T} Tavani, M., \& Arons, J.\ 1997, \apj, 477, 439 

\end{thebibliography}
\end{document}